\documentclass[%
 reprint,
 amsmath,amssymb,
 aps,
]{revtex4-2}

\usepackage{graphicx}
\usepackage{dcolumn}
\usepackage{bm}

\begin{document}

\preprint{APS/123-QED}

\title{Controllable optical sideband generation \\by breaking mechanical Parity-Time symmetry}
\author{Souvik Mondal}
 \altaffiliation{souvikjuetce95@kgpian.iitkgp.ac.in}
\author{Kapil Debnath}%
 \email{k.debnath@ece.iitkgp.ac.in}
\affiliation{%
Electronics and Electrical Communication Engineering Department, IIT Kharagpur, Kharagpur, West Bengal, 721302, India
}%
\begin{abstract}
The interaction between light and mechanical vibrations in a cavity is often exploited to produce higher order sidebands (HOS) /combs, which are used in optical communication networks, spectroscopy, and others. Although a considerable study of optomechanically induced HOS has been done, its proper control and manipulation using only continuous wave (CW) laser drive are still to be explored. Here, we employed mechanical Parity-Time ($\mathcal{PT}$) Symmetric structure with optically induced gain and loss. It has allowed us to manipulate the flow of mechanical energies (phonons) between the cavities, which has consequences on the optical response of the cavities. Based on our numerical investigations, we found that the higher order optical sidebands start to emerge by breaking $\mathcal{PT}$ symmetry. We precisely controlled the number of higher order sideband lines by adjusting the coupling rate between the cavities with fixed drive power. In addition, we observe that the Exceptional Point (EP) induces the formation of two synchronized higher order optical sideband spectra, which opens a promising EP based platform towards realization of optical readout of various mechanical synchronization phenomena, memory applications, sensing, synchronization of remote clock time and others.
\end{abstract}
\maketitle
\section{\label{sec:level1}Introduction}
Cavity optomechanics has become the subject of extensive study lately. The co-localization of the optical and the mechanical modes in an optomechanical (OM) cavity has produced many interesting outcomes ranging from classical to quantum phenomena \cite{aspelmeyer2014cavity}. One of them is the self-sustained mechanical oscillations/phonon lasing in the cavity. Once we drive  the cavity with a blue detuned CW pump laser above a certain threshold power \cite{rokhsari2006theoretical}, the mechanical vibrations in the cavity starts to amplify. The interaction of the CW pump laser with the mechanical vibrations produces Stokes and Anti-Stokes scattered light. The beating of the CW pump laser with the Stokes/Anti-Stokes scattered light results in periodic evolution of the intracavity field intensity at the mechanical resonance frequency, which again drives the mechanical vibrations coherently in the cavity. In this manner of dynamical backaction, the mechanical oscillations become self-sustained \cite{rokhsari2005radiation}. The increased driving CW laser power \cite{carmon2005temporal} or the dynamical backaction of large self sustained oscillation amplitude \cite{poot2012backaction} creates significant modulation in the intracavity field intensity evolution. As a result, we get higher order sidebands (HOS)/comb in the optical spectrum with frequency line spacing decided by the mechanical resonance frequency. Mohammad Ali Miri et al. \cite{miri2018optomechanical} showed that the formation of such HOS/comb in an OM cavity is analogous to cascaded four-wave mixing in microresonator Kerr frequency combs \cite{chembo2010modal}. The HOS can also be generated if one drives the cavity with a red detuned pump laser and an external probe field. We require an external probe as the beating of the CW pump laser with the probe at mechanical resonance frequency ensures coherent mechanical oscillations \cite{xiong2013carrier,weis2010optomechanically}.

The optical HOS/comb has many exciting applications in spectroscopy \cite{PhysRevLett.129.173204}, optical clocks \cite{diddams2002femtosecond} and others \cite{fortier201920}. So, the control and manipulation of optomechanically induced HOS/comb become crucial in a low power integrated structure. In the existing literature, the sidebands in the OM system with pump-probe drive can be manipulated in many ways. The relative phase and the power of the probe field is utilized to manipulate the optical spectrum in the cavity \cite{xiong2013carrier}. The photon tunneling rate in coupled passive optical cavities provides additional means to control the sidebands \cite{cao2016optical}. Ling Yan-He \cite{he2019parity} showed the use of active-passive coupled optical cavities in $\mathcal{PT}$ symmetry structure to enhance spectral lines. In general, the $\mathcal{PT}$ symmetry structure has been utilized in the field of optomechanics to show the low-power phonon lasing \cite{jing2014pt}, low-power chaos generation \cite{lu2015p}, enhanced OMIT \cite{li2016parity}, mass sensing \cite{djorwe2019exceptional} and others. Apart from these, hybrid atom-optomechanical system are also used to enhance and manipulate the sidebands \cite{liu2018generation}. But there exists ample scope of investigations about proper control of sidebands when it comes to driving the cavity with a CW pump laser. Conventionally, it is done by increasing CW laser drive power, but P. Djorwe et al. \cite{djorwe2020multistability} utilized the dynamical attractors to manipulate the optical combs. The quadratic optomechanical coupling \cite{zhang2014self} and the dissipation optomechanical coupling \cite{gao2021dissipation} are also being used to manipulate the sidebands.

The appearance of higher order optical sidebands depends on the amplitude of the steady mechanical oscillation \cite{marquardt2006dynamical} where the oscillations with higher amplitude results in more spectral lines \cite{poot2012backaction}. In our study, we manipulated the amplitude of the oscillations to control the sidebands using only a CW pump laser drive. This is achieved by using mechanical $\mathcal{PT}$ symmetric structure with optically induced gain and loss. The approach towards physical realization of this kind of structure can be understood in \cite{xu2015mechanical}. Based on our numerical analysis, we showed that the exception point leads to large self-sustained oscillations due to the spontaneous localization of the mechanical energy and increased optomechanical nonlinearity in the cavities \cite{zhang2015giant}. Therefore, the HOS starts to appear from EP onwards and persists in the broken $\mathcal{PT}$ symmetry phase as well. The switching among $\mathcal{PT}$ phase, EP, and broken $\mathcal{PT}$ phase is done by tuning only the mechanical coupling factor for fixed driving power. Thus, we demonstrated a different way of invoking and controlling the sidebands, contrary to conventional way of adjusting the driving power.  Practically, the mechanical coupling rate can be controlled via the piezoelectric effect \cite{okamoto2013coherent}, photothermal effect \cite{okamoto2009optical} or electrostatic force \cite{huang2013demonstration}. The generation of HOS in the system are robust against non-degenerate mechanical cavities with different resonance frequencies. This is possible because of the existence of collective phenomena in the system. In the existing literature, the collective phenomena in OM system are used in various classical and quantum synchronization purposes \cite{li2022all,sheng2020self,liao2019quantum}. Thus, we investigated EP induced synchronization between the two HOS optical spectra.

This paper is organized in the following manner. In Sec \ref{sec:level2}, we showed the mathematical model and the semiclassical dynamical equations in our coupled OM system. In Sec \ref{sec:level3}, we demonstrated mechanical $\mathcal{PT}$ symmetry and behaviour of optical spectra under different situations. In Sec \ref{sec:level4}, we studied the synchronization between the two optical HOS in the cavities. Lastly, we summarize our work in Sec \ref{sec:level5}.
\section{\label{sec:level2}Mathematical model and Dynamical equations}
The system, as shown in Fig 1 consists of two optomechanical cavities which has weak mechanical coupling.
\begin{figure}[htbp]
\includegraphics[width=\linewidth]{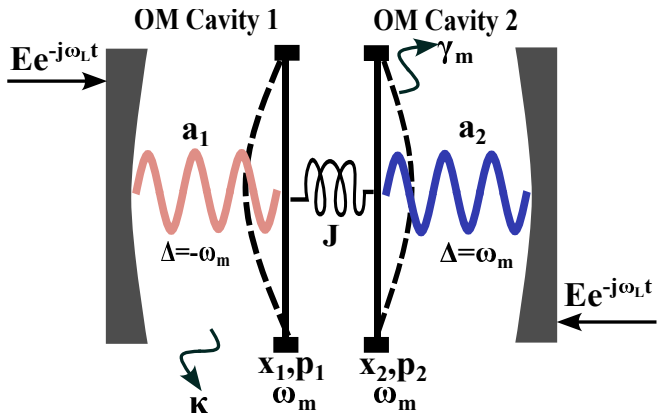}
\caption{\label{fig1}Generic setup of mechanically coupled two OM cavity}
\end{figure}
The Hamiltonian of the whole system can be represented as the sum of individual optomechanical cavity terms and the coupling terms (assuming $\hbar=1$).
\begin{table}[b]
\caption{\label{tab:table1}
Parameters in the coupled OM system}
\begin{ruledtabular}
\begin{tabular}{ll}
\textrm{Symbols} & \textrm{Meaning}\\
\hline
$E_{1,2}$ & Driving strength of the laser\\
$\kappa_{1,2}$&Optical cavity decay rates\\
$\gamma_{m_{1,2}}$&Mechanical cavity decay rates\\
$\omega_{L}$&Driving laser frequency\\
$\omega_{0_{1,2}}$&Optical cavity resonance frequencies\\
$\omega_{m_{1,2}}$ & Mechanical resonance frequencies\\
$\omega_{c_{1,2}}$ & Optical resonance frequencies\\
$\Delta_{1,2}=\omega_{L}-\omega_{c_{1,2}}$ & Detuning of the laser\\
$J$ & Mechanical coupling rate\\
$g_{0_{1,2}}$&Optomechanical coupling rate\\
\end{tabular}
\end{ruledtabular}
\end{table}
\begin{subequations}
\begin{eqnarray}
H_{OM,j}=\Delta_{j}\hat{a}^{\dagger}_{j}\hat{a}_{j}+\omega_{m_j}\hat{b}^{\dagger}_{j}\hat{b}_{j}-g_{0_{j}}\hat{a}^{\dagger}_{j}\hat{a}_{j}(\hat{b}^{\dagger}_{j}\nonumber\\+\hat{b}_{j})+E_{j}(\hat{a}^{\dagger}_{j}+\hat{a}_{j})
\\
H_{coup}=-J(\hat{b}_{j}\hat{b}^{\dagger}_{3-j}+\hat{b}^{\dagger}_{j}\hat{b}_{3-j})
\\
H_{total}=H_{OM_1}+H_{OM_2}+H_{coup}
\end{eqnarray}
\label{eq:1}
\end{subequations}
The subscript j=1,2 denotes the left and right optomechanical cavity respectively. $\hat{a}^{\dagger}_{j}(\hat{a}_{j})$ and $\hat{b}^{\dagger}_{j}(\hat{b}_{j})$ are the creation(destruction) operators of the photon and phonon respectively. The details about other parameters are mentioned in Table \ref{tab:table1}. The Hamiltonian are written in the rotating frame of the driving lasers. The dimensionless position and momentum of the mechanical cavity in terms of $\hat{b}^{\dagger}_j$ and $\hat{b}_j$ are written as $\hat{x}_j=\frac{\hat{b}_j+\hat{b}^{\dagger}_j}{\sqrt{2}}$ and $\hat{p}_j=\frac{\hat{b}_j-\hat{b}^{\dagger}_j}{i\sqrt{2}}$ respectively. Since we require the study of the dynamics of the mean values of the operators,  we can ignore the fluctuations present in the system. Using the mean field approximation ie. $\langle{\hat{a}_j\hat{x}_j}\rangle=\langle{\hat{a}_j}\rangle\langle{\hat{x}_j}\rangle$ , the dynamical equations of the mean operators($x_j=\langle{\hat{x}_j}\rangle$,$p_j=\langle{\hat{p}_j}\rangle$ and $a_j=\langle{\hat{a}_j}\rangle$) of the system can be written as
\begin{subequations}
\begin{eqnarray}
\frac{da_j}{dt}=-i(\Delta_j+g_{0_{j}}x_j)a_j-\frac{\kappa_j}{2}a_j+E_j
\\
\frac{dx_j}{dt}=\omega_{m_j}p_j
\\
\frac{dp_j}{dt}=-\omega_{m_{j}}x_j-\frac{\gamma_{m_{j}}}{2}p_j+Jx_{3-j}+g_{0_{j}}|a_j|^2
\end{eqnarray}
\label{eq:2}
\end{subequations}
The Eq (\ref{eq:2}) are coupled and non linear in nature. We evaluated the equations through numerical simulations since exact analytical derivations are difficult.We used Runge-Kutta method in obtaining the solution of Eq (\ref{eq:2}). We made certain assumptions in the parameters of the dynamical equation. The mechanical cavities are degenerate i.e. $\omega_{m_1}=\omega_{m_2}=\omega_m$. We provide identical driving power $E_1=E_2=E$ to the cavities. The decay rates and optomechanical coupling rates are identical ie. $\kappa_1=\kappa_2=\kappa$ and $g_{0_1}=g_{0_2}=g_0$. The  detunings are in resonance with the mechanical resonance frequency i.e. $-\Delta_1=\Delta_2=\omega_m$. 
\section{\label{sec:level3}$\mathcal{PT}$ Symmetry and Higher Order Sideband Generation}
We adiabatically eliminate the optical cavity modes to show mechanical $\mathcal{PT}$ symmetry. The optomechanical coupling rate $g_0$ is assumed to be weaker than the cavity decay rate $\kappa$ ie. $g_0<<\kappa$ for such condition. Including the effective loss and gain of the mechanical cavities, the effective non-Hermitian Hamiltonian is written as 
\begin{eqnarray}
H_{eff}=((\omega_m-\delta\omega_{m_1})-i\gamma_{eff_1})\hat{b}^\dagger_1\hat{b}_1\nonumber\\
+((\omega_m+\delta\omega_{m_2})+i\gamma_{eff_2})\hat{b}^\dagger_2\hat{b}_2-J(\hat{b}_1\hat{b}^\dagger_2+\hat{b}^\dagger_1\hat{b}_2)
\label{eq:3}
\end{eqnarray}
$\delta\omega_{m_j}$ is the mechanical resonance frequency shift due to the optical spring effect. The mechanical vibrations in the cavity driven by red detuned laser experience extra damping such that the effective decay rate becomes $|\gamma_{eff_1}|=\gamma_{opt_1}+\gamma_m$. The vibrations in the blue detuned driven cavity experience gain with the effective rate of $|\gamma_{eff_2}|=\gamma_{opt_2}-\gamma_m$. $\gamma_{opt_j}$ is the modifications of the decay rates due to the optomechanical coupling. At $|\Delta|=\omega_m$, $\gamma_{opt_j}$ is written as \cite{aspelmeyer2014cavity}
\begin{eqnarray}
|\gamma_{opt_j}|=\frac{64g_0^2|a_j|^2\omega^2_m}{\kappa(\kappa^2+16\omega^2_m)}
\label{eq:4}
\end{eqnarray}
Applying $\mathcal{P}$ transformation: $i\leftrightarrow i$, $\hat{b}_1\leftrightarrow-\hat{b}_2$, $\hat{b}^\dagger_1\leftrightarrow-\hat{b}^\dagger_2$ and $\mathcal{T}$ transformation: $i\leftrightarrow-i$, $\hat{b}_j\leftrightarrow\hat{b}_j$, $\hat{b}^\dagger_j\leftrightarrow\hat{b}^\dagger_j$ to Eq (\ref{eq:3}), we can show that $[H_{eff},\mathcal{PT}]=0$ under the conditions: $\delta\omega_{m_j}$ is negligible and $|\gamma_{eff_1}|=|\gamma_{eff_2}|$. The concept of $\mathcal{PT}$ symmetry is also applicable to generalised case where the gain -loss amount are unbalanced or presence of both loss-loss system. In our system, we operate in resolved sideband regime ie. $\omega_m>>\kappa$ to have minimum optical spring effect. We adjust the parameters in such a way that the total system experience net loss. It would ensure that the mechanical dynamics in $\mathcal{PT}$ phase reaches towards fixed equilibrium value and does not induce instability. With these considerations, the matrix form of Eq (\ref{eq:3}) can be written as
\begin{equation}
H_{eff}=
\begin{pmatrix}
\hat{b}^\dagger_1 & \hat{b}^\dagger_2
\end{pmatrix} 
\begin{pmatrix}
\omega_m-i\gamma_{eff_1} & -J\\
-J & \omega_m+i\gamma_{eff_2}
\end{pmatrix}
\begin{pmatrix}
\hat{b}_1\\
\hat{b}_2
\end{pmatrix}
\label{eq:5}
\end{equation}
The eigen-frequencies corresponding to the mechanical supermodes $\hat{b}_1\pm\hat{b}_2$ are written as
\begin{equation}
\omega_\pm=\omega_m-i\frac{\gamma_{eff_1}-\gamma_{eff_2}}{4}\pm\sqrt{J^2-(\frac{\gamma_{eff_1}+\gamma_{eff_2}}{4})^2}
\label{eq:6}
\end{equation}
\begin{figure}[htbp]
\includegraphics[width=\linewidth]{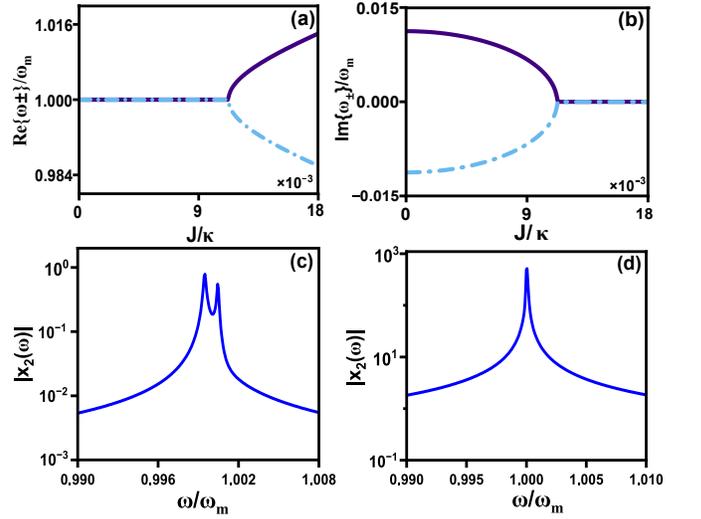}
\caption{\label{fig2} Real (a) and Imaginary (b) parts of the eigen-frequencies $\omega_{\pm}$. The exceptional point occurs at $J=0.0112\kappa$. The splitting and coalescence of the mechanical spectrum of the right cavity for the case of (c) $\mathcal{PT}$ Symmetric phase ($J=0.015\kappa$) and (d) Exceptional point}
\end{figure}
\begin{figure*}
\includegraphics[width=0.8\linewidth]{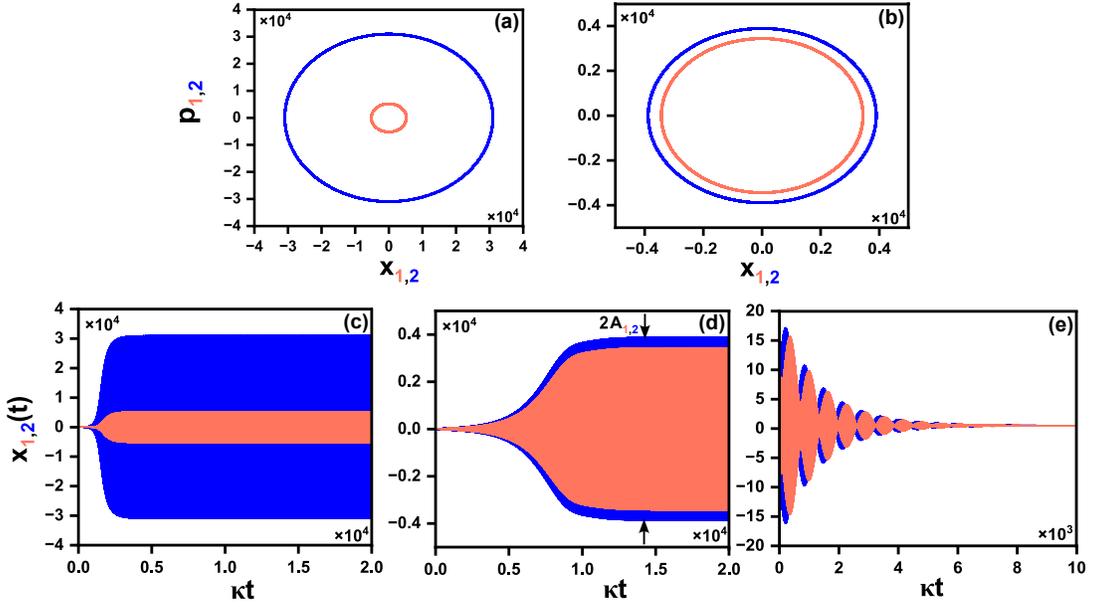}
\caption{\label{fig3} Phase portraits and dynamics of the mechanical vibrations in the two cavities. (a) and (c) show the phase portrait and mechanical oscillations dynamics in the broken $\mathcal{PT}$ phase at $J=0.002\kappa$ respectively. (b) and (d) also shows the phase portrait and mechanical oscillations dynamics at EP with $J=0.0112\kappa$ respectively. (e) shows the decaying $\mathcal{PT}$ phase dynamics at $J=0.015\kappa$. All the analysis are done under fixed $E=709\kappa$.}
\end{figure*}
In $\mathcal{PT}$ symmetry phase, the real part of eigen-frequencies of the supermodes are split by amount $\sqrt{J^2-(\frac{\gamma_{eff_1}+\gamma_{eff_2}}{4})^2}$ with a decaying component of $\frac{\gamma_{eff_1}-\gamma_{eff_2}}{4}$. In the broken $\mathcal{PT}$ phase, the splitting of the real part of eigen-frequencies vanishes. In this scenario, one of the supermodes will experience gain while other losses. The phase transition from $\mathcal{PT}$ phase to broken $\mathcal{PT}$ phase occurs at $J=\frac{\gamma_{eff_1}+\gamma_{eff_2}}{4}$ which is known as the exceptional point (EP). Now, we set the driving amplitude as $E=709\kappa$ for the rest of the study and it is low enough to avoid formation of chaos in the system. We set the intrinsic mechanical decay rate in one cavity higher than the other i.e. $\gamma_{m_1}=0.5\times10^{-2}\kappa$ and $\gamma_{m_2}=10^{-4}\kappa$. In this manner we ensure that the coupled mechanical cavities experience net loss in $\mathcal{PT}$ dynamics. The optomechanical coupling rate and the mechanical resonance frequency are set as $g_0=10^{-3}\kappa$ and $\omega_m=10\kappa$ respectively. The parameters mentioned here can be achieved in experiments too \cite{cohen2015phonon,eichenfield2009picogram}. Fig 2(a)-(b) shows the real and imaginary parts of the eigen-frequencies respectively, under the variation of mechanical coupling rate $J$. The $\mathcal{PT}$ symmetric phase is observed at high coupling rate where the Real $\omega_\pm$ are split about $\omega_m$. The EP occurs at $J=0.0112\kappa$ with the two eigen-frequencies overlapping with each other. Lowering the coupling rate will push the system into the broken $\mathcal{PT}$ phase, where one supermode experience loss while other gain. The Fig 2(c)-(d) show the nature of mechanical spectrum $|x_2(\omega)|$ of the right cavity for $\mathcal{PT}$ phase and EP respectively.
\begin{figure}[htbp]
\includegraphics[width=\linewidth]{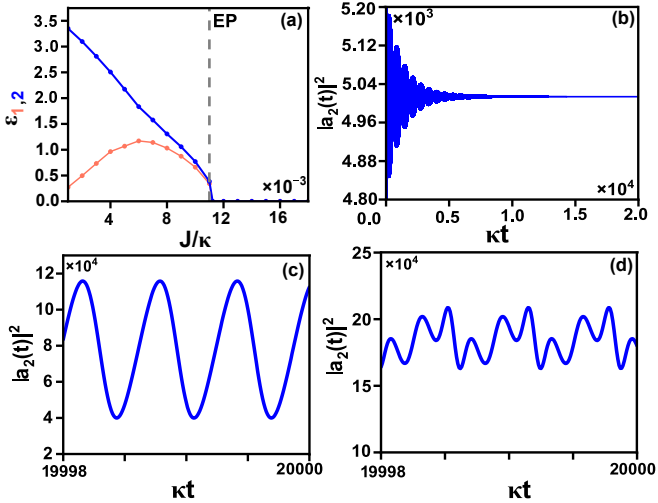}
\caption{\label{fig4} (a) Normalized self sustained oscillation amplitude ($\varepsilon_j$) vs the mechanical coupling rate $J$. The EP invokes abrupt increase of $\varepsilon_j$. (b)-(d) The steady evolution of the intracavity photon number $|a_2(t)|^2$ for $\mathcal{PT}$ phase ($J=0.015\kappa$), EP ($J=0.0112\kappa$) and broken $\mathcal{PT}$ phase ($J=0.002\kappa$) respectively.  }
\end{figure}
\begin{figure*}
\includegraphics[width=\linewidth]{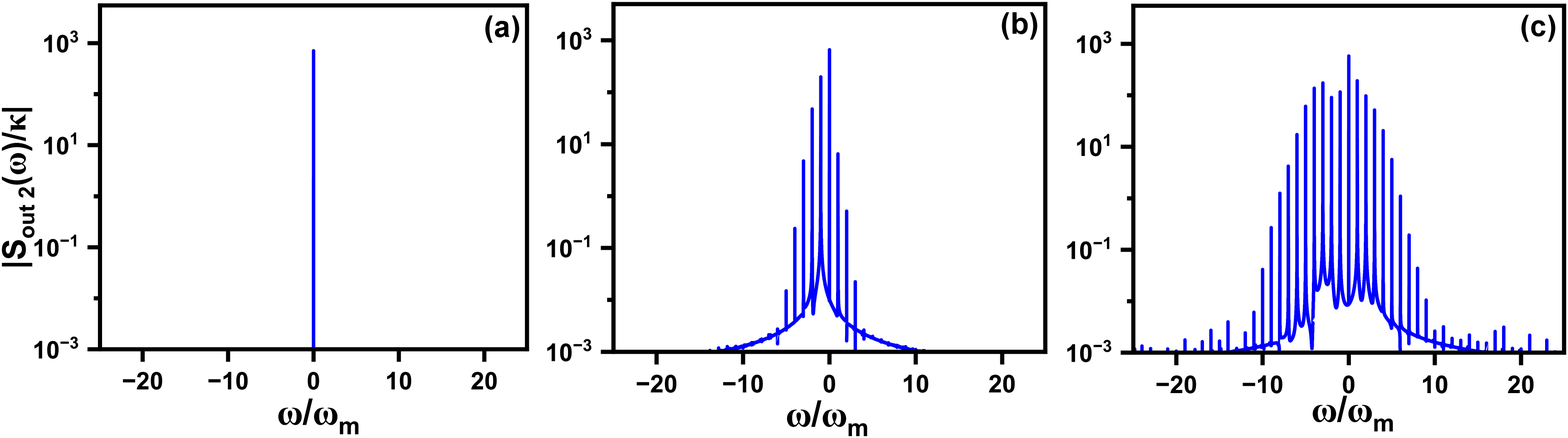}
\caption{\label{fig5} The output optical spectrum $S_{out_2}(\omega)$ of the cavity with blue detuned driving laser corresponding to the (a) $\mathcal{PT}$ phase ($J=0.015\kappa$) (b) Exceptional Point ($J=0.0112\kappa$) and (c) broken $\mathcal{PT}$ phase ($J=0.002\kappa$). }
\end{figure*}

We study the dynamics of the mechanical oscillations in different conditions by solving Eq (\ref{eq:2}) with the initial conditions of all the dynamical variables set to zero. Fig 3(e) shows the mechanical dynamics in $\mathcal{PT}$ phase for the coupling rate of $J=0.015\kappa$. We observe a periodic amplitude variation dynamics in both the cavities, which depicts strong coupling between the cavities. The mechanical dynamics has decaying envelope because of the net loss faced by the system. In the mechanical phase portrait, the system reaches towards a fixed equilibrium point as the system has decaying dynamics. Lowering the coupling rate to $J=0.0112\kappa$ takes the system at EP (Fig 3(d)). At this point, the coupling is weak enough to trigger spontaneous localization of the mechanical energy in both the cavities. It is manifested as the increase of the amplitude of oscillations in both the cavities. The amplitude will not grow forever as the optomechanical non-linearity ($g_0x_j$) becomes prominent. Thus the oscillations will eventually settle into a self sustained oscillations with amplitude $A_j$. In mechanical phase portrait diagram, the system shows limit cycle behaviour (Fig 3(b)). Lowering  the coupling rate even further at $J=0.002\kappa$, pushes the system to the broken $\mathcal{PT}$ symmetric phase as shown in Fig 3(c). Here, the dynamics show self sustained oscillations in both the cavities similar to that of EP. The mechanical oscillations in the left cavity experience optically induced loss. Lowering the coupling, reduces the supply of mechanical energy to this cavity from the optically induced gain cavity. So we observed significantly reduced steady amplitude of oscillations as shown in Fig 3(c). These are reflected in the phase portrait (Fig 3(a)) for the broken $\mathcal{PT}$ phase where the size of the limit cycle in the left cavity is significantly lesser than other.

Fig 4(a) shows the changes in the normalized amplitude $\varepsilon_j$ ($\varepsilon_j=\frac{g_0A_j}{\omega_m}$) under the variation of $J$. The amplitude are obtained by observing the mechanical dynamics. In the $\mathcal{PT}$ phase, we have $\varepsilon_j=0$. In the broken $\mathcal{PT}$ phase, we have a finite and varying $\varepsilon_j$. The magnitude of $\varepsilon_j$ is determined by the balance of mechanical losses with the driving laser power and the amount of coupling between the cavities. We observe a continuous increase of $\varepsilon_2$ towards the value $3.2$ (i.e. for the case of $J=0$) as we further reduce the coupling, while $\varepsilon_1$ decreases towards 0. For the remaining part of this section, we will only focus on the affect of vibrations on the optical response of the right cavity. Fig 4(b)-(d) shows the evolution of the intracavity photon number $|a_2(t)|^2$ in the right cavity for different conditions. In $\mathcal{PT}$ phase, $|a_2(t)|^2$ decays to steady value of around $5\times10^3$. Once EP is reached, $|a_2(t)|^2$ deviates away from steady fixed value as the finite self sustained oscillation modulates the intracavity optical field. The large self sustained mechanical oscillations (high $\varepsilon_j$) in the broken $\mathcal{PT}$ phase strongly modulates the intracavity field. Analytically, we can write the intracavity light field as infinite sum of frequency components $\pm k\omega_m$, if we assume $x_j(t)=\varepsilon_jcos(\omega_mt)$.
\begin{eqnarray}
    a_j(t)=e^{i\varepsilon_jsin(\omega_mt)} \sum_{k=-\infty}^{\infty} a_j^ke^{ik\omega_mt}
\label{eq:7}
\end{eqnarray}
In terms of $\varepsilon_j$, $a_j^k \propto J_k(-\varepsilon_j)$, where $J_k$ is the kth order first kind Bessel function \cite{marquardt2006dynamical}.

The output light field from the cavities is obtained from input-output formalism \cite{clerk2010introduction} i.e. $s_{out_j}(t)=E(t)-\kappa a_j(t)$. We obtained the output optical spectrum by taking the Fourier transform $S_{out_j}(\omega)=\int_{-\infty}^{\infty} s_{out_j}(t)e^{-i\omega t}dt$. Fig 5 shows the output optical spectrum from the right cavity under different conditions. We took Fast Fourier Transform (FFT) of the steady response of $s_{out_2}(t)$ to show the optical spectrum of the right cavity. As, we are studying in the rotating frame of driving laser, the actual spectrum is shifted by $\omega_L$. In $\mathcal{PT}$ phase, the spectrum has no sidebands (Fig 5(a)). We see a sudden transition in the nature of output spectrum at EP (Fig 5(b)) as the sidebands starts to emerge with a frequency spacing of $\omega_m$. Fig 5(c) shows significant increase of spectral lines in the broken $\mathcal{PT}$ phase. Thus, we achieved a proper control of the output spectrum from the right cavity just by varying the mechanical coupling rate. 
\section{\label{sec:level4}Synchronization between two spectra}
Till now, we observed the output optical spectrum from the blue detuned cavity with frequency spacing of $\omega_m$. The sidebands would also emerge in the red detuned cavity since there exists finite oscillation amplitude $\varepsilon_1$ from EP onwards. We show, in this section, the output HOS exists in both the optical cavities in a ``synchronized" manner.
\begin{figure*}
\includegraphics[scale=0.23]{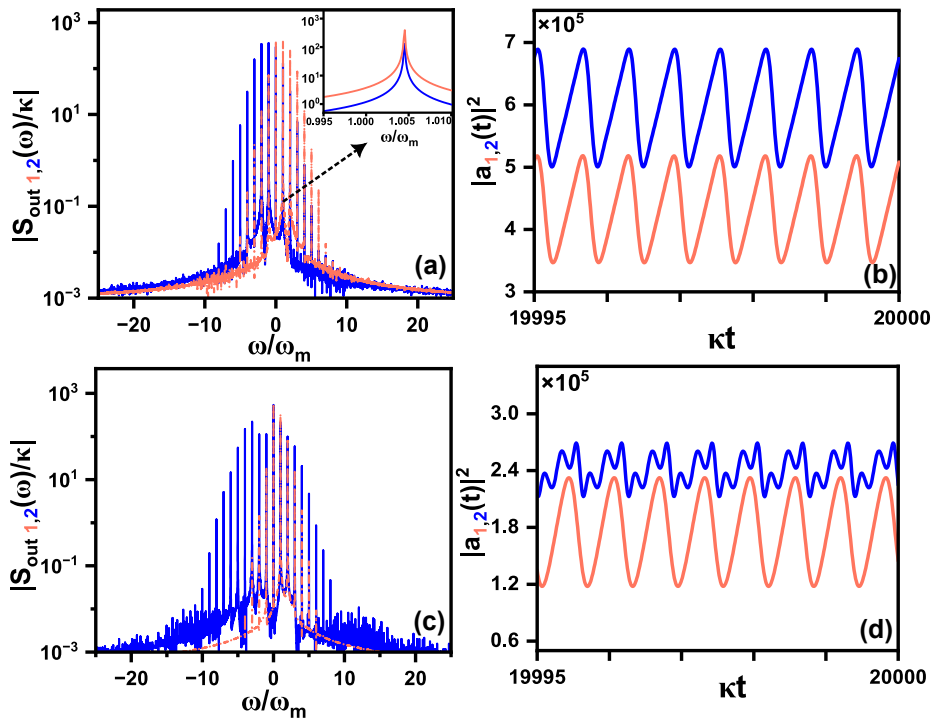}
\caption{\label{fig6} Demonstration of the collective effects in our system with a frequency mismatch of $\omega_{m_2}-\omega_{m_1}=0.0018\omega_m$. (a) and (b) Spectral profile and intracavity photon number at $J=0.07\kappa$ (synchronized case) respectively. The inset in (a) shows the 1st order spectral component coincide each other at a slight offset frequency from $\omega=\omega_m$. (c) and (d) Spectral profile and intracavity photon number at $J=0.008\kappa$ (non-synchronized case) respectively. The other parameters are same as in Fig 3.}
\end{figure*}
Here, we define synchronization in terms of (i) locking of the spectral peaks to a common frequency, if there exists any frequency mismatch and (ii) nature of spectral envelope. Since, the nature of two spectral envelope are determined by $\varepsilon_j$ (Eq (\ref{eq:7})), they become similar when the difference $|\varepsilon_2-\varepsilon_1|$ remain below a certain level.\par We set the resonance frequencies in the mechanical cavities non degenerate ie. $\omega_{m_1}\neq\omega_{m_2}$ and included the optical spring effect on the mechanical vibrations such that $\omega_{eff_{1,2}}=\omega_{m_{1,2}}\mp\delta_{\omega_{m_{1,2}}}$. This new frequency is replaced in the matrix (Eq (\ref{eq:5})) to obtain the eigenvalues of the supermodes as
\begin{eqnarray}
\omega_\pm=\frac{\omega_{eff_1}+\omega_{eff_2}}{2}-i\frac{\gamma_{eff_1}-\gamma_{eff_2}}{4}\nonumber\\
\pm\sqrt{J^2+\left(\frac{\Delta\omega}{2}+i\frac{\gamma_{eff_1}+\gamma_{eff_2}}{4}\right)^2}
\label{eq:8}
\end{eqnarray}
where $\Delta\omega=\omega_{eff_2}-\omega_{eff_1}$. In this case, the system is no longer $\mathcal{PT}$ symmetric and the EP is determined by complex interplay of coupling rate, frequency mismatch and the effective decay rates.
\begin{figure}[htbp]
\includegraphics[width=\linewidth]{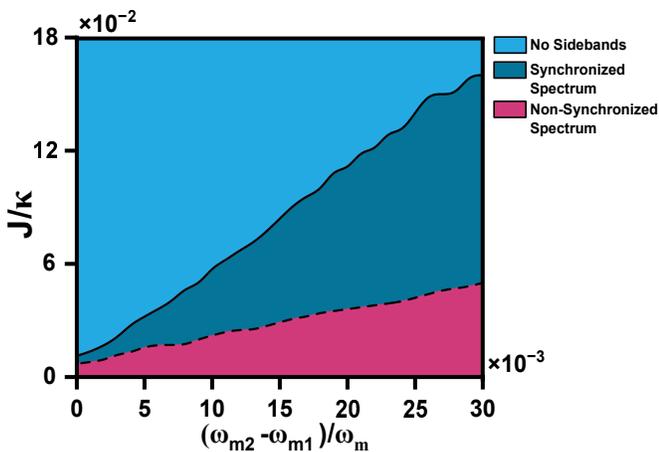}
\caption{\label{fig7} Possible regimes in our system in the parameter space of resonance frequency mismatch $(\omega_{m_2}-\omega_{m_1})/\omega_m$ and mechanical coupling rate $J$. The solid line denotes the EP and the dotted line differentiates synchronized from the non-synchronized regimes. The other parameters are same as in Fig 3.}
\end{figure}
 We know that the EP results in coalescence of the two mechanical supermodes, which leads to the mechanical vibrations in the coupled non degenerate cavities readjust itself to lock to a certain phase and amplitude, and to a common frequency \cite{djorwe2018frequency,djorwe2020self}. Such phenomenon has its impact on the output optical response of the cavities. The HOS spectra in the two cavities will emerge from EP onwards with a common frequency spacing and certain spectral envelope. Fig 6(a) shows the nature of the two output spectra. By setting the mismatch as $\omega_{m_2}-\omega_{m_1}=0.0018\omega_m$ and coupling as $J=0.07\kappa$ we observe that both the spectra has similar spectral envelope, as the difference $|\varepsilon_2-\varepsilon_1|$ is not significant. Even though the spectral envelope is similar, the lines for the left cavity is more concentrated towards $\omega>0$, while the other towards $\omega<0$. It arises because of the resolved sideband operation and the detuning of the driving laser \cite{aspelmeyer2014cavity}. Besides, both the HOS spectra has same uniform frequency spacing with the inset of Fig 6(a) showing the zoomed out version of the 1st order component around $\omega=\omega_m$. It is interesting to see the nature of evolution of the intracavity field intensity corresponding to it. We observe in Fig 6(b), that the intensity in left cavity follows the right cavity with a certain DC offset between the two. So, based on all these discussed points, we say that the two output spectra in Fig 6(a) are synchronized. Now, keeping the same frequency mismatch, but reducing the coupling to $J=0.008\kappa$, will vastly change the nature of spectrum in both the cavities (Fig 6(c)). Even though the sidebands in both the spectra locked at the common frequency spacing, the blue detuned cavity has more spectral lines than the other one since $|\varepsilon_2-\varepsilon_1|$ become significant. These are also reflected in intracavity intensity dynamics in Fig 6(d), where the evolution of $|a_2(t)|^2$ significantly becomes different than other. We say, in this scenario, that the two spectra are non synchronized.
 
We provided a region of synchronization in the parameter space of frequency mismatch and the coupling rate in Fig 7 with fixed $E=709\kappa$. We determined the region by observing the spectra, mechanical dynamics and intracavity intensity dynamics. We see that for larger mismatch, higher coupling strength is required to induce the synchronization between the spectra. Thus EP occurs at higher $J$ for high $\omega_{m_1}-\omega_{m_2}$ (solid black line in Fig 7). On the other hand, the dotted line marks the transition between the synchronized and non-synchronized spectra. In the non-synchronized regime, $\varepsilon_2 \gg \varepsilon_1$ and thus, we get results similar to the one in Fig 6 (c)-(d). But, this dotted line does not increase appreciably with the increase in mismatch. Thus, the region of synchronization becomes wider for larger mismatches. So, the process of synchronization of the spectrum is much more sensitive to EP for weak mismatches, as the synchronization region is much narrow near EP.
 
\section{\label{sec:level5}Summary}
In summary, we theoretically investigated the formation, control and synchronization of optical sideband spectra in weak mechanically coupled optomechanical cavities. The study provided rich insights behind the interplay of mechanical gain-loss structure to that of optomechanical non-linearity. The work has been done with fixed CW driving laser power (i.e. the amount of mechanical gain/loss kept fixed) with varying mechanical coupling rate. We obtained the mechanical dynamics along with the corresponding intracavity intensity dynamics under different circumstances. Consequently, we found an abrupt emergence of sidebands in the spectrum at EP by performing FFT of the steady output light field. The number of spectral lines went higher by tuning the system into deep broken $\mathcal{PT}$ phase. Thus, our scheme allows precise control of the optical spectral lines by engineering the mechanical coupling rate in a low power integrated chip-scale device. We also showed collective behaviour (synchronization) between two spectra under varied mechanical resonance frequency mismatches. Thus the study proved to be robust against the strict condition of $\mathcal{PT}$ symmetry and found the synchronization process sensitive to EP for a weak mismatch. So, the EP based generation of synchronized wideband spectra with equidistant lines provides a promising platform for optical communication technology, optical readout of mechanical synchronization process \cite{sheng2020self}, memory applications, remote clock timing synchronization \cite{deschenes2016synchronization} and others.
\bibliography{apssamp}
\end{document}